\documentclass[%
 reprint,
superscriptaddress,
amsmath,amssymb,
 aps,
prl,
longbibliography
]{revtex4-1}

\usepackage{graphicx}
\usepackage{dcolumn}
\usepackage{bm}
\usepackage{color}
\usepackage[normalem]{ulem} 
\usepackage{hyperref}

\newcommand{\YB}[1]{{\color{black} #1}}

\newcommand{\NZ}[1]{{\color{black} #1}}
\usepackage{lineno}

\begin{document}
\preprint{APS/123-QED}

\title{Search for lepton portal dark matter in \YB{the} PandaX-4T experiment}

\date{\today}

\begin{abstract}
We report a search for a lepton-portal dark matter model,  where dark matter couples to a \NZ{charged lepton in the standard model}. This simplified model naturally  \YB{leads} to photon\YB{-}mediated \YB{dark matter} interaction\YB{s} \YB{with nuclei,}  making it suitable for \NZ{direct dark matter detection experiments}. 
\YB{Matching to} the framework of \YB{non-relativistic} effective field theory \YB{for dark matter}, we report the first sensitive search  \YB{for} this model using data from \YB{the} PandaX-4T commissioning run.  \YB{Our results yield} strong constraints  \YB{on} Dirac \YB{fermion} dark matter but relatively weaker constraints on Majorana dark matter due to the suppression of effective photon interactions.  \YB{These} constraints  \YB{complement those obtained from} the collider and indirect detection experiments. 
\end{abstract}

\def\shKeyLab{School of Physics and Astronomy, Shanghai Jiao Tong University, Key Laboratory for Particle Astrophysics and Cosmology (MoE), Shanghai Key Laboratory for Particle Physics and Cosmology, Shanghai 200240, China}
\def\scKeyLab{Jinping Deep Underground Frontier Science and Dark Matter Key Laboratory of Sichuan Province, Liangshan 615000, China}
\def\BUAA{School of Physics, Beihang University, Beijing 102206, China}
\def\BUAACenter{Peng Huanwu Collaborative Center for Research and Education, Beihang University, Beijing 100191, China}
\def\BUAALab{Beijing Key Laboratory of Advanced Nuclear Materials and Physics, Beihang University, Beijing 102206, China}
\def\SCNT{Southern Center for Nuclear-Science Theory (SCNT), Institute of Modern Physics, Chinese Academy of Sciences, Huizhou 516000, China}
\def\USTClab{State Key Laboratory of Particle Detection and Electronics, University of Science and Technology of China, Hefei 230026, China}
\def\USTCdep{Department of Modern Physics, University of Science and Technology of China, Hefei 230026, China}
\def\BUAALab{International Research Center for Nuclei and Particles in the Cosmos \& Beijing Key Laboratory of Advanced Nuclear Materials and Physics, Beihang University, Beijing 100191, China}
\def\pku{School of Physics, Peking University, Beijing 100871, China}
\def\YaLongSD{Yalong River Hydropower Development Company, Ltd., 288 Shuanglin Road, Chengdu 610051, China}
\def\IAP{Shanghai Institute of Applied Physics, Chinese Academy of Sciences, Shanghai 201800, China}
\def\CHEPpku{Center for High Energy Physics, Peking University, Beijing 100871, China}
\def\SDUdep{Research Center for Particle Science and Technology, Institute of Frontier and Interdisciplinary Science, Shandong University, Qingdao 266237, China}
\def\SDUlab{Key Laboratory of Particle Physics and Particle Irradiation of Ministry of Education, Shandong University, Qingdao 266237, China}
\def\UMD{Department of Physics, University of Maryland, College Park, Maryland 20742, USA}
\def\TDLee{New Cornerstone Science Laboratory, Tsung-Dao Lee Institute, Shanghai Jiao Tong University, Shanghai 201210, China}
\def\MESJTU{School of Mechanical Engineering, Shanghai Jiao Tong University, Shanghai 200240, China}
\def\SYU{School of Physics, Sun Yat-Sen University, Guangzhou 510275, China}
\def\SYUSFI{Sino-French Institute of Nuclear Engineering and Technology, Sun Yat-Sen University, Zhuhai 519082, China}
\def\NKU{School of Physics, Nankai University, Tianjin 300071, China}
\def\YTU{Department of Physics, Yantai University, Yantai 264005, China}
\def\FDU{Key Laboratory of Nuclear Physics and Ion-beam Application (MOE), Institute of Modern Physics, Fudan University, Shanghai 200433, China}
\def\USST{School of Medical Instrument and Food Engineering, University of Shanghai for Science and Technology, Shanghai 200093, China}
\def\SJTUSC{Shanghai Jiao Tong University Sichuan Research Institute, Chengdu 610213, China}
\def\SPEIT{SJTU Paris Elite Institute of Technology, Shanghai Jiao Tong University, Shanghai 200240, China}
\def\NNU{School of Physics and Technology, Nanjing Normal University, Nanjing 210023, China}
\def\SYSUzhuhai{School of Physics and Astronomy, Sun Yat-Sen University, Zhuhai 519082, China}
\def\CDUT{College of Nuclear Technology and Automation Engineering, Chengdu University of Technology, Chengdu 610059, China}
\def\WISC{Department of Physics, University of Wisconsin-Madison, WI 53706, USA}

\affiliation{\TDLee}
\author{Xuyang Ning}\affiliation{\shKeyLab}
\author{Zihao Bo}\affiliation{\shKeyLab}
\author{Wei Chen}\affiliation{\shKeyLab}
\author{Xun Chen}\affiliation{\TDLee}\affiliation{\shKeyLab}\affiliation{\SJTUSC}\affiliation{\scKeyLab}
\author{Yunhua Chen}\affiliation{\YaLongSD}\affiliation{\scKeyLab}
\author{Zhaokan Cheng}\affiliation{\SYUSFI}
\author{Xiangyi Cui}\affiliation{\TDLee}
\author{Yingjie Fan}\affiliation{\YTU}
\author{Deqing Fang}\affiliation{\FDU}
\author{Zhixing Gao}\affiliation{\shKeyLab}
% ...
\collaboration{PandaX Collaboration}
\author{Yang Bai}\email[Corresponding author: ]{yangbai@physics.wisc.edu}\affiliation{\WISC}
\noaffiliation

\maketitle

Plenty of astronomical and cosmological observations indicate the existence of dark matter (DM)~\cite{Bertone:2004pz}, but the nature of DM remains unknown. \NZ{It is usually assumed that DM is made up of non-baryonic, neutral and fundamental particles. However, there are still some possibilities that dark matter possesses minute electromagnetic couplings with photons which evades the current experimental constraints.} Recently, we reported a search of DM's electromagnetic properties with the PandaX-4T experiment~\cite{PandaX:2023toi}, which uses the effective field theory (EFT) approach and provides constraints on individual photon-mediated operators. It turns out that the most stringent constraint on the charge radius square of DM is stronger than that of the neutrinos by approximately four orders of magnitude. 
\NZ{To understand the origin of these possible electromagnetic properties, we need to go beyond the EFT theory to the ultraviolet (UV)-complete theories. A generic UV-complete model, running down to low-energy scales, will generally give rise to multiple types of EFT operators simultaneously. In addition, DM searches with collider experiments, especially those conducted at the Large Hadron Collider (LHC), are performed at significant higher energy scale where the validity of EFT may break down. Therefore,  a UV-complete model is necessary to combine the results from direct detection, indirect detection and collider searches. }


In this work, our focus is on one specific realization of photon-mediated interaction: the lepton-portal DM (LPDM) model~\cite{leptonPortal1,leptonPortal2,leptonP_thorough}. \NZ{The general construction involves assuming a portal coupling between the dark sector and standard model (SM) leptons.}
Usually, a neutral DM candidate $\chi$ and a charged scalar mediator with a heavier mass are assumed in the dark sector. This model is gaining increasing attentions  due to the additional loop contributions from the dark sector and its potential to explain the muon anomalous magnetic moment~\cite{leptonP_thorough,Bai_g-2,Kawa_g-2,Liu:2021mhn}. In \NZ{direct DM detection experiments}, this model can naturally generate the photon-mediated loop process between DM candidates and the target nuclei. This process can be expressed within the framework of DM EFT approach by matching it to the relevant effective operators. In this letter, we perform a sensitive search for the LPDM  using the EFT treatment \NZ{with} the data from PandaX-4T commissioning run for the first time.


 \YB{In the} LPDM,  \YB{within} a minimal  setup, two new particles in the dark sector are introduced. The lighter one is the DM candidate, \YB{while} the other  serves as a charged mediator \YB{facilitating interactions} between DM and SM leptons. \YB{In the minimal model,}  the dark sector  \YB{exclusively} couples to the right-handed SM lepton~\cite{leptonPortal2}. Depending on the nature of the DM candidate, the forms of photon-mediated interactions and \YB{their} corresponding signatures in direct detection \YB{experiments}  \YB{vary}. In this work,  \YB{our} focus \YB{is} on the  \YB{scenario} where DM is a fermion, either Dirac or Majorana, and the interaction is mediated by an $SU(2)_{L}$ singlet particle. 

The corresponding Lagrangian for a fermion DM in LPDM is:
\begin{equation}
\mathcal{L}_{\rm LPDM} \supset \lambda_{i}\phi_{i}\overline{\chi}_{L}e^{i}_{R} + h.c.
\end{equation}
The free parameters include the charged mediator $\phi$ mass $m_\phi$, DM mass $m_\chi$, and the coupling $\lambda_{i}$ to \YB{a specific} lepton flavor $i$.  To avoid any flavor violating process from the dark sector,  \YB{we assume that the coupling occurs exclusively with a single SM lepton in the mass eigenstate at any given time.}
As an example,  we take muon as the only portal\YB{-linked}  \YB{lepton} particle  \YB{in this work}. \NZ{This scenario offers a possibility to address phenomena such as the muon magnetic anomaly~\cite{Muong-2:2023cdq}.}

\begin{figure}[h]
  \centering
  \includegraphics[width=\columnwidth]{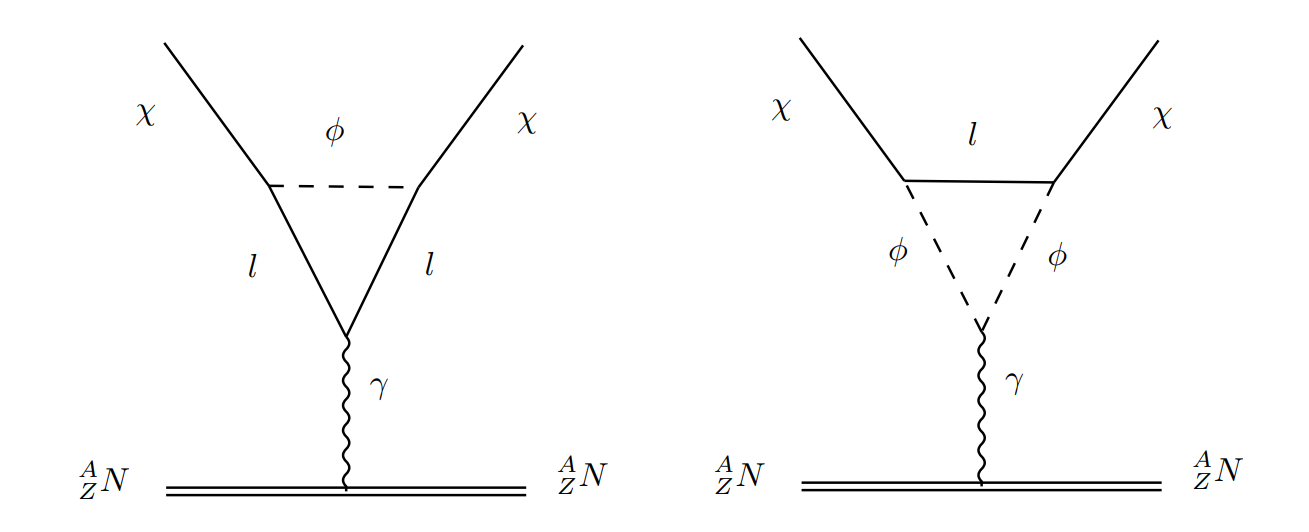}
  \caption{Feynman diagrams contributing to \YB{the} \YB{photon-mediated} interaction between DM and \YB{the} nucleus at \YB{the} loop level.}
  \label{fig:feyn}
\end{figure}

In direct detection, since \YB{the} dark sector  interacts \YB{exclusively} with \YB{charged} leptons, the leading contribution  \YB{to} DM scattering off a nucleus is a loop process mediated by photon. The relevant Feynman diagrams \NZ{induced by photon} are shown in Fig.~\ref{fig:feyn}.
Additional diagrams induced by $Z$ or Higgs boson are also possible, but they are subdominant compared to the electromagnetic moments and can be neglected safely~\cite{chargedMediator}. 

 \YB{The photon-mediated} process contains multipole effective interactions,  
\begin{equation}
\begin{aligned}
\mathcal{L}_{\rm LPDM} =&\; \mathcal{L}_{\rm CR}+\mathcal{L}_{\rm MD}+\mathcal{L}_{\rm A}+\mathcal{L}_{\rm ED} \\
=&\; b_{\chi}\overline{\chi}\gamma^{\mu}\chi\partial^{\nu}F_{\mu\nu}
                            +\frac{\mu_{\chi}}{2}\overline{\chi}\sigma^{\mu\nu}\chi F_{\mu\nu}\\
                            &\; + a_{\chi}\overline{\chi}\gamma^{\mu}\gamma^{5}\chi\partial^{\nu}F_{\mu\nu}
                            +i\frac{d_{\chi}}{2}\overline{\chi}\sigma^{\mu\nu}\gamma^{5} \chi F_{\mu\nu} ~,
\end{aligned}
\end{equation}
where $b_{\chi}$, $\mu_{\chi}$, $a_{\chi}$ and $d_{\chi}$ are the Wilson coefficients corresponding to  \YB{DM} charge radius (CR), magnetic dipole (MD), anapole (A), and electric dipole (ED) moments. The values of these coefficients are determined by the LPDM parameters, and  \YB{their} renormalization group evolution  \YB{results in} a logarithmic enhancement, as shown in the Appendix. 
\NZ{These relativistic effective operators lead to interactions between DM and nuclei mediated by photons. }
For the Dirac \YB{fermion} DM case, interactions  \YB{through} the charge radius $b_{\chi}$ and the magnetic dipole $\mu_{\chi}$ are particularly  \YB{significant}, \YB{while} the anapole \YB{moment} $a_{\chi}$ is suppressed\YB{,} and the electric dipole \YB{moment} $d_\chi$ vanishes in this simple setup. For  \YB{Majorana fermion} DM, only the anapole \YB{moment} $a_\chi$  is non-vanishing. As seen here, this UV\YB{-}complete model \YB{leads to various} types of relativistic operators and \YB{their mixing}, \YB{making it challenging to directly translate constraints from a single operator.} 

\begin{figure}[h]
  \centering
    \includegraphics[width=\columnwidth]{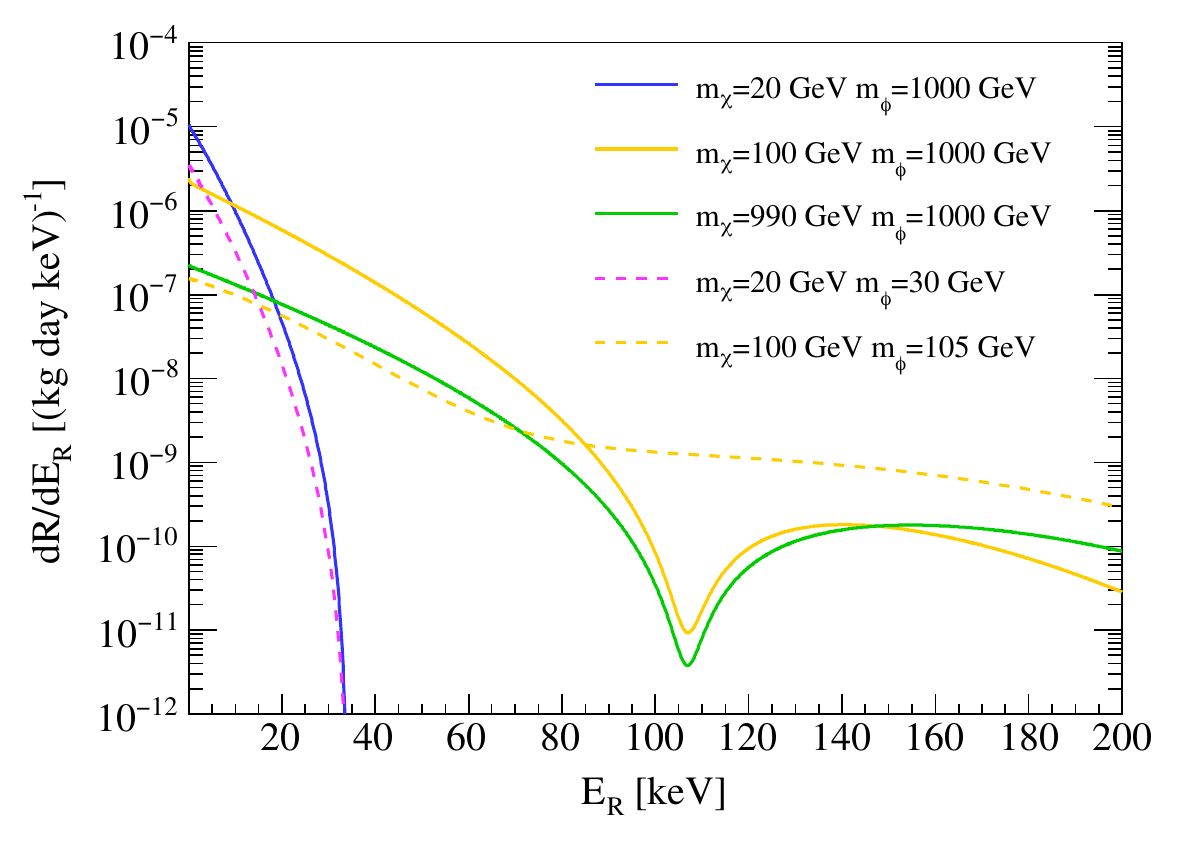}
  \caption{Differential event rate \NZ{as a function of nuclear recoil energy $E_R$} for lepton portal DM scattering off a xenon nucleus,  \YB{assuming a} coupling $\lambda=1$. Different DM mass\YB{es} and mediator mass\YB{es} are shown in  \YB{various} colors. Solid lines are for the Dirac \YB{fermion} DM,  \YB{while} dashed lines are for the Majorana \YB{fermion} DM. 
  }
  \label{fig:spec_lp}
\end{figure}

\NZ{In the non-relativistic limit for direct detection, these photon-mediated interactions are further matched to linear combinations of non-relativistic EFT (NREFT) operators~\cite{DMff2}.} \YB{These NREFT operators are} constructed based on \YB{parameters such as} the momentum exchange $q$, DM velocity $v$, DM spin $S_\chi$\YB{,} and nucleon spin $S_N$~\cite{pole_dic,Fitzpatrick:2012ib,DirecDM}, shown in Appendix for details. 
 \YB{Utilizing} the corresponding Wilson coefficients and relevant effective operators, the differential event rate in the xenon detector can be derived using DMFormFactor~\cite{DMff1,DMff2}, which  \YB{incorporates} the xenon nuclear response functions from \YB{comprehensive} full-basis shell-model calculations. The nuclear recoil (NR) energy spectrum of \YB{a} xenon nucleus in LPDM is shown in Fig.~\ref{fig:spec_lp}.  \YB{In the case of}  Majorana fermion DM, the DM scattering cross section is suppressed by the DM velocity or \NZ{NR energy} $E_{R}$,  \YB{resulting in} a small \YB{predicted} signal rate. 

\YB{The} PandaX-4T experiment, located in the B2 hall of \YB{the} China Jinping Underground Laboratory (CJPL), operates a dual-phase time projection chamber (TPC) with 3.7 tonne liquid xenon in the sensitive volume, aiming to explore new physics  \YB{related to both} DM and neutrino\YB{s}. Prompt scintillation light ($S1$) and delayed electroluminescence photons from ionized electrons ($S2$) of a scattering event are collected by two arrays of 3-inch photomultiplier tubes (PMTs). The 3-dimensional scattering position is reconstructed  \YB{using} the time difference between $S1$ and $S2$ in \YB{the} z\YB{-}direction,  \YB{along with the} $S2$ top PMT patterns in \YB{the} x-y direction. The deposited energy  \YB{is} reconstructed  \YB{based on} the charge of $S1$ and $S2$. The signal response model in PandaX-4T follows the standard NEST v2.2.1 construction~\cite{NESTv2, szydagis2021review}, with parameters fitted from the \NZ{low-energy calibration data which were taken at the end of run}. 

\begin{figure}[h]
  \centering
    \includegraphics[width=\columnwidth]{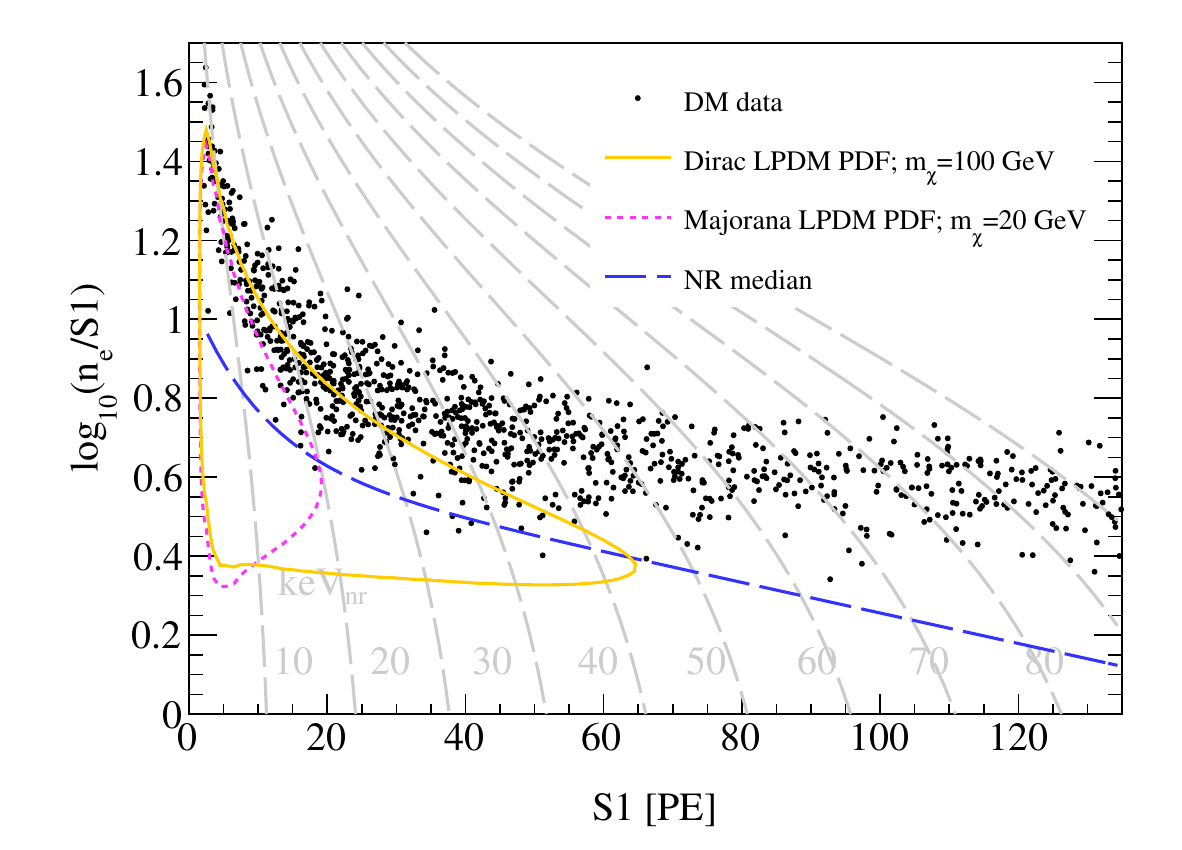}
  \caption{The distribution of the final candidate events in PandaX-4T experiment in ${\rm log_{10}}(n_e/S1)$ vs. $S1$, where $n_e$ is the number of ionized electrons.
  The green (magenta) line is the $2\sigma$ contour of the expected distribution of 100~GeV$/c^2$ Dirac LPDM (20~GeV$/c^2$ Majorana LPDM) with mediator mass $m_\phi=1~{\rm TeV}/c^2$ ($30~{\rm GeV}/c^2$) and coupling strength $\lambda=1$. The dashed blue line is the fitted median of the nuclear recoil events from neutron calibration data.
  }
  \label{fig:band}
\end{figure}

The data used to search for the LPDM correspond to \YB{an} 86.0 live-day exposure  \YB{during} the PandaX-4T commissioning run. The event selection criteria 
follow the \NZ{analysis for WIMP searches}~\cite{PandaX-4T:2021bab}, with the region of interest (ROI) defined as $S1$ \YB{ranging} from 2 to 135 \NZ{photoelectrons (PEs)} \YB{and} raw $S2$ from 80 to 20,000 PEs.  \YB{A} total \YB{of} 1058 events are identified  \YB{within} the ROI, \NZ{as shown in Fig~\ref{fig:band}}. \NZ{The expected number of events from the background} is $1054\pm 39$,  \YB{which includes sources such as} tritium, ${}^{85}$Kr, radon, material radioactivity, surface, ${}^{136}$Xe, neutrons, neutrinos and accidental $S1-S2$ coincidence events, as summarized in Ref.~\cite{PandaX-4T:2021bab}. \YB{We construct} a two-sided profile likelihood ratio  to test the signal hypothesis,  \YB{using} the standard unbinned likelihood~\cite{Baxter:2021pqo}  \YB{along} with Gaussian penalty terms to account for the uncertainty  \YB{associated with} each nuisance parameter.  \YB{Probability density functions \NZ{(PDFs)} for both the background components and the signal are generated based on the signal response model in $S1$ versus $S2_b$ (bottom-only $S2$) space.}

\begin{figure}[h]
  \centering
  \includegraphics[width=\columnwidth]{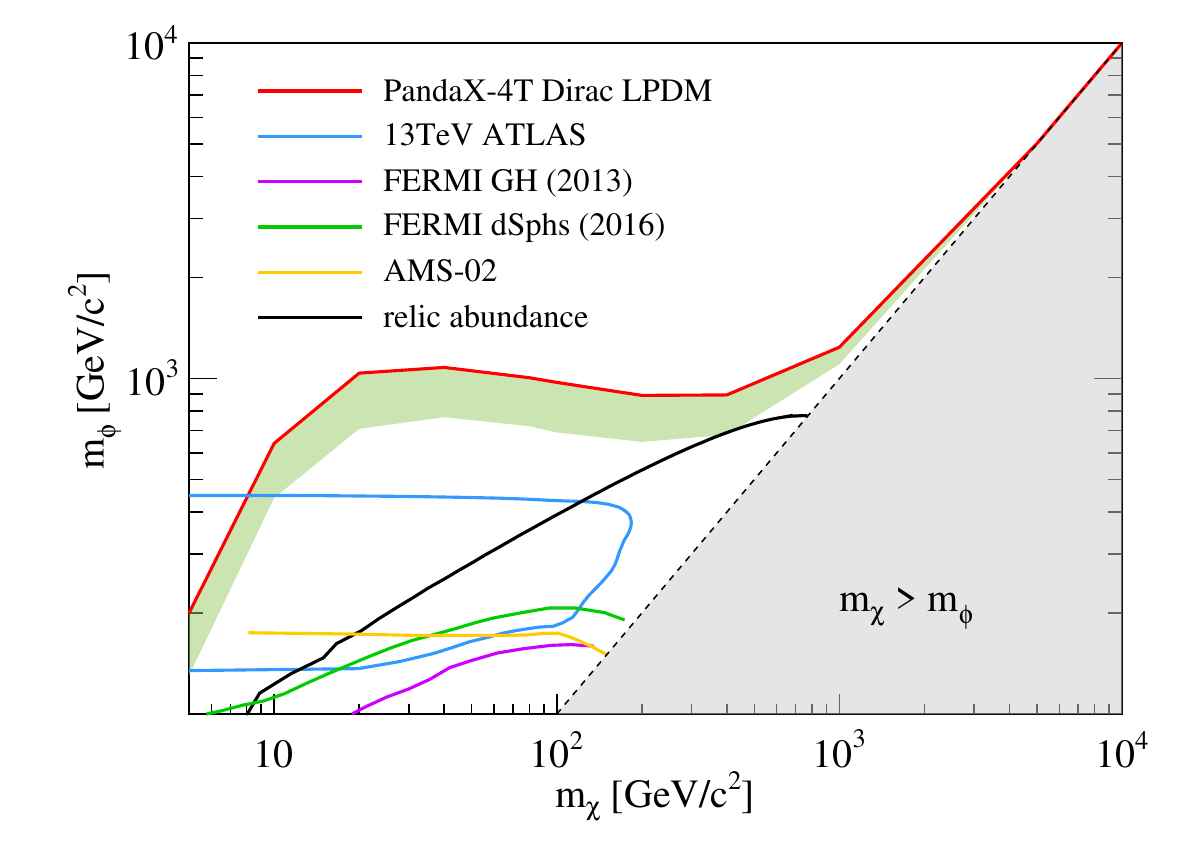}
  \includegraphics[width=\columnwidth]{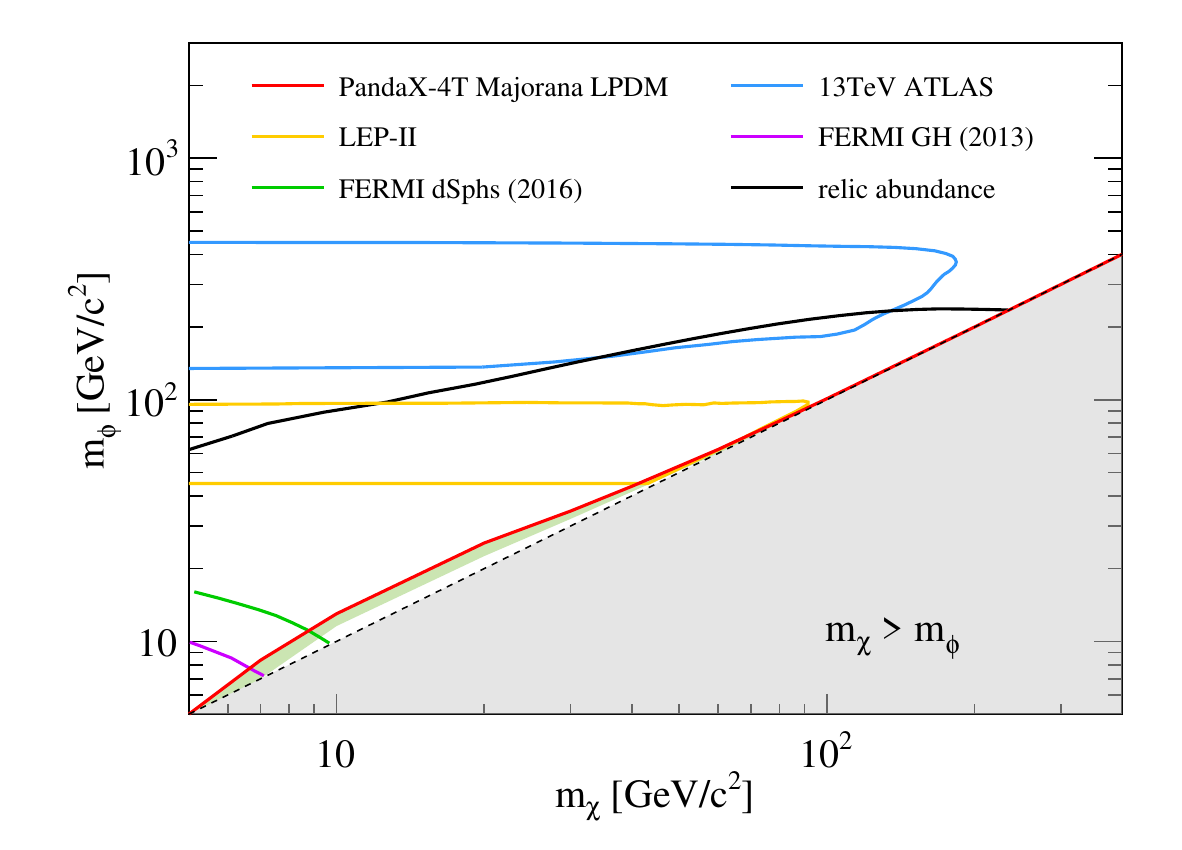}
  \caption{Constraints on \YB{the} mediator mass as a function of dark matter mass in the LPDM for both Dirac \YB{fermion} DM (upper \YB{panel}) and Majorana \YB{fermion} DM (lower \YB{panel}). The green band represents $\pm1\sigma$ sensitivity band. \YB{A coupling strength of $\lambda=1$ is assumed.} \YB{The} collider results are taken from \NZ{the search for the right-handed slepton signals by the ATLAS experiment}~\cite{ATLAS} and LEP-II experiment~\cite{LEP}. Indirect detection bounds are  \YB{sourced} from~\cite{leptonPortal2,leptonPortal1}. \YB{Additionally,} the parameter \YB{region} for LPDM \YB{that}  \YB{satisfies} the thermal relic density~\cite{leptonPortal2} is  \YB{indicated},  \YB{predominantly governed} by the $\phi$-mediated tree-level annihilation process. 
  }
  \label{fig:limit_lp}
\end{figure}

The constraints on  LPDM are shown in Fig~\ref{fig:limit_lp}, as a function of the DM mass $m_\chi$ and the mediator mass $m_\phi$ 
\YB{with a} coupling strength $\lambda=1$.  \YB{It's important to note that, for the stability of DM,} only the region  \YB{where} $m_\phi > m_\chi$ is allowed. For the Dirac fermion DM case,  LPDM  in collider experiments exhibits signature analogous to the pair production of right-handed sleptons from the Minimal Supersymmetric Standard Model (MSSM)~\cite{Cahill-Rowley:2012asv,Cahill-Rowley:2014boa}. \NZ{The latest constraints from the ATLAS slepton searches with two-lepton plus large missing transverse energy final state in $\rm 139~fb^{-1}$ of data are also included~\cite{ATLAS}}. Direct detection  \YB{provides} compelling result\YB{s}  \YB{across most of} the \YB{plotted} mass range. \YB{However,} for DM mass\YB{es} less than 8 GeV$/c^2$,  \YB{collider experiments present stronger constraints due to the decrease in efficiency for lower recoil energy regions in direct detection.} \NZ{Indirect detection experiments also contribute constraints due to the DM direct annihilation into leptons in LPDM.} This includes measurements such as the positron fraction measurement  \YB{by} AMS-02~\cite{AMS}, as well as the indirect limits on $\left \langle \sigma v \right \rangle$ derived from FERMI observations of the Galactic Halo~\cite{fermi_GH} and Milky Way Dwarf~\cite{fermi_dsf}, as depicted in Fig~\ref{fig:limit_lp}. 
 \YB{It's worth noting that our results exclude the parameter combinations of Dirac DM that would satisfy the thermal relic density.}
 \YB{In the case of}  Majorana \YB{fermion} DM, \YB{the constraints from direct detection are weaker due to velocity or recoil energy suppression, which leaves a substantial parameter space open to address the muon magnetic moment anomaly}.

In summary,  \YB{utilizing} data from \YB{the} PandaX-4T commissioning run, \YB{we conduct a search for} a UV-complete lepton portal dark matter model\YB{,} which introduces photon-mediated \YB{interactions between DM and nucleons.}  Constraints on the mediator and DM masses are derived,  \YB{assuming a} muon-only portal  \YB{with} $\lambda=1$.  \YB{Our findings provide} the strongest constraints for the Dirac \YB{fermion} DM and  \YB{encompass} the largest parameter space for Majorana \YB{fermion} DM  \YB{with} DM mass\YB{es} \YB{exceeding}  $10~{\rm GeV}/c^2$.  
\YB{These results suggest} that\YB{,} for the UV complete model, direct detection  \YB{offers} promising constraint and can  complement  collider or indirect detection experiment\YB{s,}  \YB{given their distinct advantages in different regions}. The PandaX-4T experiment continues  \YB{to collect} physics data and will  \YB{further explore these uncharted parameter spaces}.


\section{Acknowledgement}
 
We thank Wick Haxton, Zuowei Liu and Natsumi Nagata for helpful discussions. 
This project is supported in part by grants from National Science Foundation of China (Nos. 12090060, 12090061, 12325505, U23B2070), a grant from the Ministry of Science and Technology of China (Nos. 2023YFA1606200, 2023YFA1606201), and by Office of Science and Technology, Shanghai Municipal Government (grant No. 22JC1410100, 21TQ1400218). We thank for the support by the Fundamental Research Funds for the Central Universities. We also thank the sponsorship from the Chinese Academy of Sciences Center for Excellence in Particle Physics (CCEPP), Hongwen Foundation in Hong Kong, New Cornerstone Science Foundation, Tencent Foundation in China, and Yangyang Development Fund. Finally, we thank the CJPL administration and the Yalong River Hydropower Development Company Ltd. for indispensable logistical support and other help. 

\bibliographystyle{apsrev4-1}
\bibliography{refs.bib}
\appendix
\section{Appendix}
The photon-mediated relativistic effective operators can be matched to the linear combination of non-relativistic DM-nucleons operators~\cite{PandaX:2023toi, DMff2, relic_new,pole_dic,Fitzpatrick:2012ib,DirecDM,leptonPortal1}: 
\begin{equation}
\begin{aligned}
\mathcal{L}_{\rm MD} &= \frac{\mu_{\chi}}{2}\overline{\chi}\sigma^{\mu\nu}\chi F_{\mu\nu} \\
\rightarrow  ~
&\mathcal{O}_{\rm MD} = \mu_{\chi}\frac{8\pi^{2}}{2e}\left[-\frac{\alpha}{2\pi}Q_{N}\left(\frac{\mathcal{O}_{1}}{m_{\chi}}+4\frac{m_{N}}{q^{2}}\mathcal{O}_{5}\right)\right.\\
&\left.\quad \quad \quad-\frac{\alpha}{\pi}\frac{g_{N}}{m_N}\left(\mathcal{O}_{4}-\frac{m^{2}_{N}}{q^{2}}\mathcal{O}_{6}\right) \right],\\
\mathcal{L}_{\rm ED} &= i\frac{d_{\chi}}{2}\overline{\chi}\sigma^{\mu\nu} \gamma^{5}\chi F_{\mu\nu} \\ 
\rightarrow ~
&\mathcal{O}_{\rm ED} = d_{\chi}\frac{8\pi^{2}}{2e}\frac{2\alpha}{\pi}\frac{m_N}{q^2}Q_{N}\mathcal{O}_{11},\\
\mathcal{L}_{\rm A} &= a_{\chi}\overline{\chi}\gamma^{\mu}\gamma^{5}\chi\partial^{\nu}F_{\mu\nu}\\
\rightarrow ~
&\mathcal{O}_{\rm A} = ea_{\chi}\left(2Q_{N}\mathcal{O}_{8}-g_{N}\mathcal{O}_{9}\right),\\
\mathcal{L}_{\rm CR} &= b_{\chi}\overline{\chi}\gamma^{\mu}\chi\partial^{\nu}F_{\mu\nu}\\ \rightarrow ~
&\mathcal{O}_{\rm CR} = eb_{\chi}Q_{N}\mathcal{O}_{1},\\
\end{aligned}
\end{equation}

where $Q_{p} = 1$, $Q_{n}=0$, $g_{p}=5.59$, $g_{n}=-3.83$, $\alpha=1/137$. The relevant non-relativistic EFT operators $\mathcal{O}_i$ are~\cite{DMff2}:
\begin{equation}
\begin{aligned}
\mathcal{O}_{1} &=1 \,,\\
\mathcal{O}_{4} &=\Vec{S}_{\chi} \cdot \Vec{S}_{N} \,,\\
\mathcal{O}_{5} &=i \Vec{S}_{\chi} \cdot \left(\frac{\Vec{q}}{m_{N}} \times \Vec{v}^{\perp} \right) \,,\\
\mathcal{O}_{6} &=\left( \Vec{S}_{\chi} \cdot \frac{\Vec{q}}{m_{N}} \right)
             \left( \Vec{S}_{N} \cdot \frac{\Vec{q}}{m_{N}} \right) \,,\\
\mathcal{O}_{8} &=\Vec{S}_{\chi} \cdot \Vec{v}^{\perp} \,,\\
\mathcal{O}_{9} &=i \Vec{S}_{\chi} \cdot \left( \Vec{S}_{N} \times \frac{\Vec{q}}{m_{N}} \right) \,,\\
\mathcal{O}_{11} &=i \Vec{S}_{\chi} \cdot \frac{\Vec{q}}{m_{N}} \,,\\
\end{aligned}
\end{equation}
where $\Vec{S}_{\chi}$ is the DM spin, $\Vec{S}_{N}$ is the nucleon spin, $m_N$ is the nucleon mass, $\Vec{q}$ is the momentum exchange, and $\Vec{v}^{\perp}$ is the component of the relative velocity $\Vec{v}_{\rm rel}$ orthogonal to $\Vec{q}$. 

For the minimal lepton portal dark matter model with \YB{the} assumption that DM is a fermion, the  \YB{matched} Wilson coeﬀicients for Dirac \YB{fermion} DM are~\cite{leptonP_thorough}:
\begin{equation}
\begin{aligned}
\left(b_{\chi}\right)_{\ell} &=\frac{e Q_{\ell}\left|\lambda\right|^{2}}{16 \pi^{2} m_{\chi}^{2}} \hat{b}_{\chi}\left(\mu, \epsilon\right), \\
\left(\mu_{\chi}\right)_{\ell} &=\frac{e Q_{\ell}\left|\lambda\right|^{2}}{16 \pi^{2} m_{\chi}} \hat{\mu}_{\chi}\left(\mu, \epsilon\right), \\
\left(a_{\chi}\right)_{\ell} &=\frac{e Q_{\ell}\left|\lambda\right|^{2}}{16 \pi^{2} m_{\chi}^{2}} \hat{a}_{\chi}\left(\mu, \epsilon\right), \\
\left(d_{\chi}\right)_{\ell} &=0 ,
\end{aligned}
\end{equation}
where
\begin{equation}
\begin{aligned}
\hat{b}_{\chi}(\mu, \epsilon) &=-\frac{1}{24}\left[(8 \mu-8 \epsilon+1) \log \left(\frac{\epsilon}{\mu}\right)\right.\\
&\left.+4\left(4+\frac{\mu+3 \epsilon-1}{\Delta}\right)\right.\\
&\left.+\frac{2}{\Delta^{3 / 2}}\left[8 \Delta^{2}+(9 \mu+7 \epsilon-5) \Delta\right.\right. \\
&\left.\left.-4 \epsilon(3 \mu+\epsilon-1)\right] \tanh ^{-1}\left(\frac{\Delta^{1 / 2}}{\mu+\epsilon-1}\right)\right] \,, \\
\hat{\mu}_{\chi}(\mu, \epsilon)&=-\frac{1}{2}\left[\frac{1}{2}(\epsilon-\mu) \log \left(\frac{\epsilon}{\mu}\right)-1\right. \\
&\left.-\frac{\Delta+\mu+\epsilon-1}{\Delta^{1 / 2}} \tanh ^{-1}\left(\frac{\Delta^{1 / 2}}{\mu+\epsilon-1}\right)\right] \,, \\
\hat{a}_{\chi}(\mu, \epsilon)&=\frac{1}{12}\left[\frac{3}{2} \log \left(\frac{\epsilon}{\mu}\right)\right.\\
&\left.+\frac{3 \mu-3 \epsilon+1}{\Delta^{1 / 2}} \tanh ^{-1}\left(\frac{\Delta^{1 / 2}}{\mu+\epsilon-1}\right)\right] \,, \\
\end{aligned}
\end{equation}
with $ \mu \equiv m_{\phi}^{2} / m_{\chi}^{2}, \ \epsilon \equiv m_{\mu}^{2} / m_{\chi}^{2},\ \Delta \equiv \mu^{2}+(\epsilon-1)^{2}-2 \mu(\epsilon+1)$ and the charge of muon $Q_l=-1$.

For Majorana \YB{fermion} DM, the differential rate is obtained by replacing $a_{\chi}$ to $2a_{\chi}$, while \YB{keeping} others zero.

\end{document}